\documentclass[twocolumn,aps,pra,floatfix,showpacs,noshowkeys,epsfig,graphics]{revtex4}%
\usepackage{graphicx}
\usepackage{amsmath}
\usepackage{amsfonts}
\usepackage{amssymb}
\newcommand{\bk}{{\mathbf k}}

\newcommand{\cd}{c^\dag}

\newcommand{\br}{{\mathbf r}}
\newcommand{\up}{\uparrow}
\newcommand{\down}{\downarrow}
\newcommand{\ket}[1]{|#1\rangle}
\newcommand{\bra}[1]{\langle#1|}

\newcommand{\hpsi}{\hat{\psi}}
\newcommand{\hpsid}{\hat{\psi}^\dag}
\begin{document}
\title{Entanglement of electron spins in superconductors}
\author{Sangchul Oh}\email{scoh@kias.re.kr}
\author{Jaewan Kim}\email{jaewan@kias.re.kr}
\affiliation{School of Computational Sciences, 
    Korea Institute for Advanced Study, Seoul 130-012, Korea}
\date{\today}
\begin{abstract}
We investigate entanglement of two electron spins forming Cooper pairs in 
an $s$-wave superconductor. The two-electron space-spin density matrix is 
obtained from the BCS ground state using a two-particle Green's function. 
It is demonstrated that a two spin state is not given by a spin singlet
state but by a Werner state. It is found that the entanglement 
length, within which two spins are entangled, is not the order of the 
coherence length but the order of the Fermi wave length.
\end{abstract}
\maketitle

Entanglement, referring to the nonlocal quantum correlation between 
subsystems, is one of the key resources in quantum teleportation, 
quantum communication, and quantum computation~\cite{Nielsen01}. 
A study of entanglement in many-body systems is of importance for not only 
its application to quantum information processing but also giving us new 
insights on its relevant physics. For example, entanglement in many spin 
systems has been investigated in connection with quantum phase 
transition~\cite{Verstraete04}. 
For a non-interacting electron gas, the entanglement length within which 
two electron spins are entangled is the order of the Fermi wave 
length $\lambda_F = 2\pi/k_F$~\cite{Vedral03,Oh04}. 

A solid state entangler, analogous to the parametric down conversion 
producing a pair of entangled photons in quantum optics, is a device 
to generate a pair of entangled electrons in controlled way. It is of 
interest for the realization of scalable solid-state quantum computers. 
There have been various proposals to create entangled pairs in solid 
state systems; spin-entanglement via a quantum dot~\cite{Saraga03,Oliver02} 
or via a magnetic impurity~\cite{Costa01}, generation of entangled 
electron spins by extracting a Cooper pair out of 
a superconductor~\cite{Recher01,Recher02,Recher03,Bena02,Lesovik01,
Chtchelkatchev02,Bouchiat03}, entangled electron-hole pairs in 
a degenerate electron gas~\cite{Beenakker03,Beenakker04a}, 
two particle orbital entanglement~\cite{Samuelsson03,Samuelsson04},
and entangled spins in electron gases due to exchange 
interaction~\cite{Vedral03,Oh04,Lebedev03}. 
Usually entangling process takes two steps, generation of 
an entangled electron pair via some kind of interaction and separation 
of them from each other.

A Cooper pair of a BCS superconductor is composed of two electrons with 
opposite momenta and a spin singlet state, $\bk\uparrow$ and 
$-\bk\downarrow$~\cite{Bardeen57}. Since the size of a Cooper pair is 
the order of the coherence length $\xi$ ($\sim 10^{-4}$ cm), entanglement 
of electron spins may survive within that scale. This means the entanglement 
length may be about the coherence length. If one can 
extract a Cooper pair and separate two electrons from each other, 
the superconductor may be a good natural resource of entangled spin states.
In Refs.~\cite{Recher01,Recher02,Recher03,Bena02,Lesovik01, Chtchelkatchev02,
Bouchiat03} it is implicitly assumed that the distance between two tunnel 
junctions attached to the superconductor should be less than the coherence 
length $\xi$. Also it is unclear whether a spin state of a tunneled electron 
pair is a spin singlet state (one of four Bell states) or a mixed state. 

In this Letter we address this problem and find the two-spin state 
of the BCS ground state is not given not by a Bell state but by 
a Werner state. We investigate entanglement of the two-spin state 
as a function of the relative distance between two electrons. 
Surprisingly, we find the entanglement length is not the order of 
the coherence length $\xi$ but the order of the Fermi wave length 
$\lambda_F$.
 
Let us start with the pairing Hamiltonian of the BCS theory~\cite{Bardeen57,Schrieffer}
\begin{equation}
H = \sum_{\bk s} \epsilon_{\bk}\, c_{\bk s}^\dag c_{\bk s}
  + \sum_{\bk,\bk'} V_{\bk\bk'}\, c_{\bk\up}^\dag c_{-\bk\down}^\dag c_{-\bk'\down}
     c_{\bk'\up} \,, 
\label{Eq:BCS_Hamiltonian}
\end{equation}
where $c_{\bk s}^\dag$ is a creation operator for electrons of wave vector 
$\bk$ and $z$-component of spin $s$. The normalized BCS ground state 
of Eq.~(\ref{Eq:BCS_Hamiltonian}) is given by
\begin{eqnarray}
\ket{\psi_0} 
= \prod_{\bk}\bigl( u_\bk + v_\bk c_{\bk\up}^\dag c_{-\bk\down}^\dag \bigr) \ket{0} \,,
\end{eqnarray}
where coefficients $u_{\bf k}$ and $v_{\bf k}$ are written by 
\begin{subequations}
\begin{eqnarray}
u_{\mathbf k}^2 &=& \frac{1}{2} 
                \left ( 1 + \frac{\epsilon_{\mathbf k} -\mu}{E_{\mathbf k}}\right)\,, \\
v_{\mathbf k}^2 &=& \frac{1}{2} 
                \left ( 1 - \frac{\epsilon_{\mathbf k} -\mu}{E_{\mathbf k}}\right)\,.
\end{eqnarray}
\end{subequations}
Here $E_{\mathbf k} = \sqrt{(\epsilon_{\mathbf k} - \mu)^2 + \Delta_{\mathbf k}^2}$ is
the excitation energy of a quasi-particle of wave vector $\bk$ and 
$\Delta_{\mathbf k}$ the superconducting gap.

In order to investigate the entanglement of two-electron spins forming Cooper pairs,
we introduce  the two-electron space-spin density matrix in the second quantization 
\begin{eqnarray}
\rho^{(2)}(x_1,x_2;x_1',x_2')
= \frac{1}{2} \langle \hat{\psi}^{\dag}(x_2') \hat{\psi}^{\dag}(x_1')
                             \hat{\psi}(x_1)\hat{\psi}(x_2) \rangle \,,
\end{eqnarray}
where $\langle\cdots\rangle = \langle\psi_0|\cdots|\psi_0\rangle$ at zero temperature and
$x =({\mathbf r},s)$.
The two-particle Green's function is defined by
\begin{equation}
G(1,2;1',2') 
= - \langle T[\hpsi_H(1)\hpsi_H(2) 
              \hpsi_H^\dag(2')\hpsi_{H}^\dag(1') ] \rangle\,,
\end{equation}
where 1 refers to $x_1t_1$, $\hpsi_{H}^\dag$ is the creation operator for electrons
in the Heisenberg representation, and  $T$ the time-ordering operator. 
The relation between the two-electron space-spin 
density matrix and the two-particle Green's function reads
\begin{eqnarray}
\rho^{(2)}(x_1,x_2;x'_1,x_2')
= -\frac{1}{2}G(x_1t_1, x_2t_2 ; x_1't_1^{+}, x_2't_2^{+})\,,
\end{eqnarray}
where $t^{+}$ denotes time infinitesimally later than $t$.

The two-particle Green's function for the superconducting state 
can be factored into single-particle Green's functions~\cite{Fetter,Abrikosov}
\begin{align}
G(1,2;1',2') &= G(1,1')G(2,2') - G(1,2')G(2,1') \nonumber\\
             & -  F(1,2)F^{\dag}(1',2') \,.
\label{Eq:two_Green}
\end{align}
The single-particle Green's function is given by
\begin{subequations}
\label{Eq:single_Green}
\begin{align}
G(1,1')
&\equiv -i\langle T[\psi_{H}(x_1t_1)\psi_{H}^\dag(x'_1t'_1)]\rangle  \\
&= \delta_{s_1s_1'}\,G(\br_1 t_1,\br'_1t'_1) \,,
\end{align}
\end{subequations}
where its spin dependence for a non-magnetic system becomes a unit matrix $\delta_{s_1s_1'}$. 
The anomalous Green's function is written by
\begin{subequations}
\label{Eq:anomalous_Green}
\begin{align}
F^{\dag}(1,2) 
&\equiv -i\langle\, T[\psi_{H}^{\dag}(x_1t_1)\psi_{H}^{\dag}(x_2t_2)]\,\rangle  \\
&= I_{s_1s_2} F^{\dag}(\br_1t_1,\br_2t_2) \,,
\end{align}
where its spin dependence is given by an antisymmetric matrix 
\begin{eqnarray}
I_{ss'} 
= \begin{bmatrix}
  0 & 1 \\
  -1 & 0 
  \end{bmatrix} = i\sigma_y \,.
\end{eqnarray}
\end{subequations}
Also notice that $F(1,2) = I_{s_1s_2}F(\br_1t_1,\br_2t_2)$.

\begin{figure}[htbp]
\includegraphics[scale=1.0,angle=0]{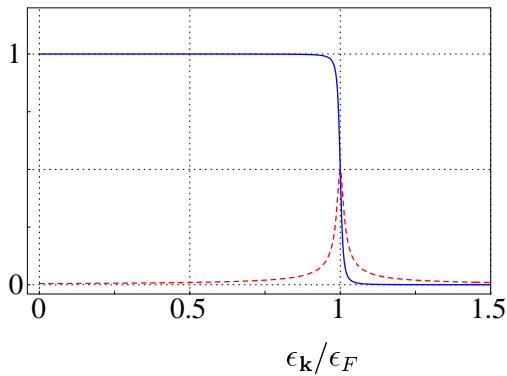}
\caption{(color online) $v_{\bk}^2$ (solid line) and $u_{\bk}v_{\bk}$ (dotted line) 
as a function of $\epsilon_{\bf k}/\epsilon_F$. In plotting $\Delta/\epsilon_F$ is taken to 
be $1/1000$.}
\label{Fig:vk2}
\end{figure}

Since the system is translational-invariant and the Hamiltonian of 
the system is time-independent, the Green's functions $G$ and $F$ depend 
only on the relative coordinates $\mathbf{r}_1 - \mathbf{r}_1'$ and 
the time difference $t_1 - t_1'$. In the limit that $t_1'$ goes to $t_1$, 
the spatial part of $G$ becomes
\begin{subequations}
\begin{align}
iG(\br)
   &= \bra{\psi_0} \hpsid_{\sigma}(\br_1)\hpsi_{\sigma}(\br_1')\ket{\psi_0} \\
   &= \frac{1}{V}\sum_{\bk} v_{\bk}^2 \, e^{i\bk\cdot\br} \,,
\end{align}
\end{subequations}
where 
\begin{eqnarray*}
\hpsid_{s}(\br) = \frac{1}{\sqrt{V}}\sum_{\bk} \cd_{\bk s} e^{i\bk\cdot\br} \,.
\end{eqnarray*}
is the field operator in Schr\"odinger representation.
In the continuum limit one obtains 
\begin{align}
iG(r) = \frac{1}{2\pi^2r}\frac{m}{\hbar^2} \int_0^\infty v_k^2\sin(kr) k\, dk\,.
\label{Eq:iG}
\end{align}
Due to the similarity between $v_{\bf k}^2$ of the BCS ground state
and the Fermi function of an ideal electron gas at the critical temperature $T_c$, 
as shown in Fig.~\ref{Fig:vk2}, Eq.~(\ref{Eq:iG}) has a similar form of Eq.~(17) 
in Ref.~\cite{Oh04}.
The electron density $n \equiv N/V$ can be calculated by $iG(0) = n/2$.
Similarly, the spatial part of $F^\dag$ is given by
\begin{subequations}
\begin{align}
iF^\dag(\br_1-\br_2) &= \bra{\psi_0} \hpsid_{\up}(\br_1)\hpsid_{\down}(\br_2)\ket{\psi_0} \\
             &= \frac{1}{V}\sum_{\bk} v_{\bk} u_{\bk} e^{i\bk\cdot(\br_1-\br_2)}\,.
\end{align}
\end{subequations}
Note that $F^\dag(\br_1-\br_2) = \left[F(\br_1 -\br_2)\right]^*$.
In the continuum limit one has
\begin{align}
iF(r)&=\frac{1}{\pi^2 r}\int_{\cal R}\frac{\sin(kr)k}{\sqrt{(\xi/\Delta)^2 + 1}}\, dk 
\label{Eq:iF}
\end{align}
where $\xi_\bk\equiv \epsilon_{\bk} -\mu$, 
$\Delta_{\bk} = \Delta\,\theta(\hbar\omega_D - |\xi_\bk|)$ with a step function $\theta(x)$,
and the integration over $k$ should be done on the range such that 
$-\hbar\omega_D\le\xi_\bk\le\hbar\omega_D$~\cite{Bardeen57,Schrieffer,Fetter}. 
Here $\omega_D$ is the Debye frequency.
Thus Eq.~(\ref{Eq:iF}) becomes approximately
\begin{align}
iF(r) &\approx N(0)\Delta \frac{\sin(k_Fr)}{k_Fr} K_0(\textstyle{\frac{r}{\pi\xi_0}}) 
\end{align}
where $N(0)$ is the density of states for one spin projection at the  Fermi surface,
given by~\cite{Fetter}
\begin{equation}
N(0) = \frac{1}{2\pi^2}\Bigl[k^2\frac{dk}{d\epsilon_k}\Bigr]_{\epsilon_k=\epsilon_F}
     = \frac{mk_F}{2\pi^2\hbar^2}\,,
\end{equation}
and $K_0(y)$ is a Bessel function of order 0
\begin{eqnarray}
K_0(y) = \int_0^\infty dt\,\frac{\cos(yt)}{\sqrt{1+t^2}} 
      \approx \sqrt{\frac{\pi}{2y}} e^{-y} \,.
\end{eqnarray}
From Eqs. (\ref{Eq:two_Green}), (\ref{Eq:single_Green}) and (\ref{Eq:anomalous_Green}), 
we have the two-electron space-spin density matrix
\begin{align}
\rho^{(2)}_{s_1,s_2;s_1',s_2'}&(\br_1,\br_2;\br_1'\br_2')  \nonumber\\
  =-\frac{1}{2}\bigl[\>
            &\>\delta_{s_1s_1'}\delta_{s_2s_2'}\, G(\br_1 -\br_1')G(\br_2 -\br_2') \nonumber\\
           -&\>\delta_{s_1s_2'}\delta_{s_2s_1'}\, G(\br_1 -\br_2')G(\br_2 -\br_1') \nonumber\\
           -&\>I_{s_1s_2}I_{s_1',s_2'}\, F(\br_1 -\br_2)F^*(\br_1' -\br_2')\> \bigr]\,.
\label{Eq:ss_density}
\end{align}
Eq.~(\ref{Eq:ss_density}) has the same form of the two-electron space-spin 
density matrix for a non-interacting electron gas, Eq.~(8) in Ref.~\cite{Oh04}, 
except the last anomalous term. 
In the limit that $|\br_i-\br_i'|\to \infty$ for $i =1,2$,
one has
\begin{align}
\rho^{(2)}_{s_1,s_2;s_1',s_2'}&(\br_1,\br_2;\br_1'\br_2') \nonumber\\
\to\> &\>\frac{1}{2} I_{s_1s_2}I_{s_1',s_2'}\, F(\br_1 -\br_2)F^*(\br_1' -\br_2')
\end{align}
which shows the off-diagonal long range order of a superconductor~\cite{Yang62}.

Let us consider the case $\br_1=\br_1'$, $\br_2 = \br_2'$, which is equivalent to take 
only the diagonal elements of the space density matrix. For a solid state entangler 
that produce entangled spins out of a superconductor, two leads are attached to two 
tunneling points $\br_1$ and $\br_2$ on the superconductor. This implies one electron is 
located at $\br_1$ and the other at $\br_2$. The two spin state depending on 
the relative distance $\br\equiv\br_1-\br_2$ is given by 
\begin{align}
\rho^{(2)}_{s_1,s_2;s_1',s_2'}(r) 
  =- \frac{1}{2}\Bigl[\,&
     \delta_{s_1,s_1'} \delta_{s_2,s_2'} G^2(0)
   - \delta_{s_1,s_2'}\delta_{s_2,s_1'}G^2(r) \nonumber \\
  &- I_{s_1,s_2}I_{s_1',s_2'}|F(r)|^2 \,\Bigr] \,.
\label{Eq:two_spin_state}
\end{align}
Let us define functions $g(r) \equiv G(r)/G(0) = 2iG(r)/n$ and
$f(r) \equiv F(r)/G(0)$. Eq.~(\ref{Eq:two_spin_state}) becomes
\begin{align}
\rho^{(2)}_{s_1,s_2;s_1',s_2'}(r) 
  = \frac{n^2{\cal N}}{8} \rho_{12}\,,
\end{align}
where the two-spin state $\rho_{12}$ of the BCS ground state 
with normalization factor ${\cal N}\equiv 4 - 2g^2 + 2f^2$ is given 
by
\begin{align}
\label{Eq:Werner}
\rho_{12} &= \frac{1}{\cal N}\begin{bmatrix}
               1 -g^2& 0          & 0          & 0 \\
               0      & 1 + f^2    & -g^2 - f^2 & 0 \\
               0      & -g^2 - f^2 & 1 + f^2    & 0 \\
               0      &  0         & 0          & 1 -g^2
               \end{bmatrix}\,.
\end{align}
We find that the two-spin state $\rho_{12}$ is not a spin singlet state but 
a Werner state characterized by a single parameter $p$ $(0\le p \le 1)$ 
\begin{align}
\rho_{12} = (1-p)\frac{\mathbf{I}}{4} + p|\Psi^{(-)}\rangle\langle \Psi^{(-)}|\,,
\end{align}
where $\mathbf{I}$ is a $4\times 4$ unit matrix and 
$\ket{\Psi^{(-)}} = \frac{1}{\sqrt{2}}\left(\ket{\up\down} - \ket{\down\up}\right)$.
The parameter $p$ is given by a function of $f$ and $g$
\begin{eqnarray}
p = \frac{f^2+g^2}{2+f^2 -g^2} \,.
\end{eqnarray}
For a non-interacting electron gas, the two spin-state is also given by 
a Werner state and can be obtained from Eq.~(\ref{Eq:Werner}) by putting 
$f=0$~\cite{Oh04}.

\begin{figure}
\includegraphics[scale=1.0,angle=0]{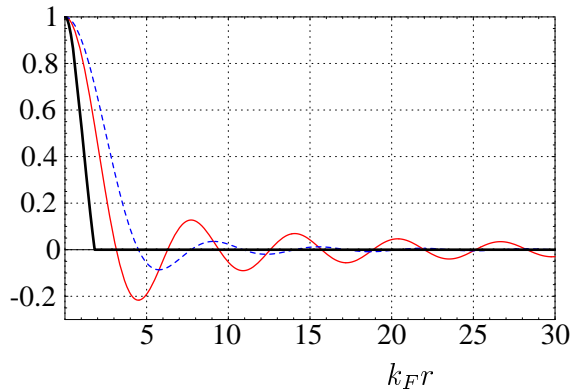}
\caption{(color online) $\widetilde{F}$ (solid line), $g$ (dotted line), 
         $C$ (thick solid line) as a function of $k_Fr$. The plotting parameters 
          are given in the text.}
\label{Fig:f_k}
\end{figure}

The properties of a Werner state are well known~\cite{Oh04,Werner89}. According 
to the Peres-Horodecki separability criterion~\cite{Peres96,Horodecki96}, 
a Werner state is entangled for $p > 1/3$. 
Thus $\rho_{12}$ is entangled for $f^2 + 2g^2 >1$.
If $g^2=0$, $\rho_{12}$ is entangled for $f^2 \ge 1$. As shown later, $f^2$ is always 
much less than 1. The concurrence for $\rho_{12}$, one of entanglement 
measures, is calculated as~\cite{Wootters98,Oh04}
\begin{align}
C = \max\left\{0,\frac{3p -1}{2}\right\}\;.
\end{align}
Fig.~\ref{Fig:f_k} depicts the concurrence as a function of $k_Fr$.  
This implies that the two spin state $\rho_{12}$ of the BCS ground state is not entangled 
if the relative distance $r$ between two tunneling points on the superconductor is 
larger than the Fermi wave length $\lambda_F$ and even though less than the coherence 
length $\xi$. 

Let us calculate the magnitude of $f(r)$ which is rewritten as
\begin{align}
   f(r) = \frac{F(r)}{F(0)}\frac{F(0)}{G(0)}  \,,
\end{align}
where $\widetilde{F}(r) \equiv F(r)/F(0)$ is always less then 1 except at the origin 
as shown in Fig.~\ref{Fig:f_k}.
On the other hand we obtain
\begin{subequations}
\label{Eq:f_g}
\begin{align}
\frac{F(0)}{G(0)} &\approx \frac{2}{n} N(0)\Delta \ln\frac{2\hbar\omega_D}{\Delta}\\
&= \frac{3}{2}\frac{\Delta}{\epsilon_F}\ln\frac{2\hbar\omega_D}{\Delta} \,.
\end{align}
\end{subequations}
In the weak coupling limit~\cite{Bardeen57,Schrieffer,Fetter}, 
$\Delta \ll \hbar\omega_D \ll \epsilon_F$, 
the value of Eq.~(\ref{Eq:f_g}) is very small. For example, for the values 
$\Delta \approx 1\> \text{\rm meV}$, $\hbar\omega_D \approx 100\>\text{\rm meV}$, 
and $\epsilon_F\approx 1\>\text{\rm eV}$, we have $F(0)/G(0) \sim 10^{-2}$.
This means $f^2$ is very small compared with $g^2$. Thus we demonstrate
that two spins are not entangled for $\lambda_F < r < \xi$.

Our results could be explained as follows. A superconductor has many Cooper 
pairs. Although the spin state of each Cooper pair is a Bell state, the spin 
correlation between two points $\br_1$ and $\br_2$ is due to many Cooper 
pairs not a single Cooper pair. Thus for solid state entangler which
generate entangled spins from a superconductor, one needs to extract 
a single Cooper pair or to confirm that two electrons are tunneled out from 
a single Cooper pair.

In conclusion, we obtained the two-electron reduced density matrix of 
the BCS ground state based on the Green's function method. 
We investigated entanglement of two electron-spins forming 
the Cooper pair in BCS superconductors. 
It has been found that the two-spin density 
matrix for a given relative distance between two electrons is given by a Werner state 
not by a Bell state. Also the entanglement length is not the order of 
the coherence length $\xi$ but the order of the Fermi wave length $\lambda_F$.\\


J.K. was supported by Korean Research Foundation Grant KRF-2002-070-C00029.
S.O. was partially supported by R\&D Program for Fusion Strategy of Advanced 
Technologies of Ministry of Science and Technology of Korean Government.

\end{document}